\documentclass[fleqn,12pt]{JHEP3}
\usepackage{amsfonts,epsfig}
\usepackage{latexsym,amssymb}

\newcommand{\ft}[2]{{\textstyle\frac{#1}{#2}}}
\def\tilde{\widetilde}

\def\1bar{1\hskip -.275cm -}
\def\2bar{2\hskip -.275cm -}
\def\3bar{3\hskip -.275cm -}

\newsavebox{\uuunit}

\makeatletter \@addtoreset{equation}{section} \makeatother

\def\bfone{\relax{\rm 1\kern-.35em 1}}

\def\bfone{\relax{\rm 1\kern-.35em 1}}
 
\newcommand{\bbZ}{{\mathbb{Z}}}

\providecommand{\mathring}[1]{{\stackrel{\mbox{\tiny o}}{#1} }}
\newcommand{\mathon}{\mathversion{bold}}
\newcommand{\mathoff}{\mathversion{normal}}

\newcommand{\be}{\begin{equation}}
\newcommand{\ee}{\end{equation}}
\newcommand{\bea}{\begin{eqnarray}}
\newcommand{\eea}{\end{eqnarray}}
\newcommand{\ben}{\begin{displaymath}}
\newcommand{\een}{\end{displaymath}}

\newcommand{\nn}{\nonumber}

\title{ String Theory on Dp-plane waves}
\author{ Jose F. Morales \\
{\em Institute for Theoretical Physics} \,and\, {\em Spinoza Institute},\\
{\em Utrecht University, Postbus 80.195, 3508 TD Utrecht, The Netherlands}}
\abstract{
We study the spectrum of solvable string models
on plane waves descending from non-conformal Dp-brane geometries.
We mainly focus on S-dual F1/D1-waves in type IIB and
type I/heterotic 10D superstrings.
 We derive the Kaluza-Klein spectrum of ${\cal N}=1,2$
10D supergravities on D1/F1-waves.
  We compute helicity supertraces counting
multiplicities and ${\cal R}$-charges of string excitations in the
plane wave geometry.
The results are compared against the expectations
coming from gauge/supergravity descriptions.
  In the type I case, the Klein, Annulus and Moebius one-loop
amplitudes are computed for ten-dimensional
D1-waves. We test the consistency
of the open string descendant by showing that after modular
transformations to the closed string channel,
the three amplitudes combine themselves to reconstruct a complete square
$(|B\rangle+|C\rangle)^2$. Tadpole conditions are also discussed. }
\begin{document}

\section{Introduction}

According to \cite{maldacena}, gauge theories at large N
can often be rephrased in terms of holographically related
string theories on $AdS\times S$ spaces pierced by
RR fluxes. The typical example relates
${\cal N}=4$ SYM to type IIB string theory on $AdS_5\times S^5$.
   Superstring theories on such backgrounds are still
far from being tractable, obstructing a more quantitative study
of the correspondence beyond the classical supergravity level.
An unexpected handle for this physics comes from
\cite{Berenstein:2002jq}, where
a solvable string theory
was proposed as an effective description of the bulk physics
in the nearby of null geodesics of $AdS_5\times S^5$.
 The geometry is defined by a Penrose limit along a generic
null geodesic of  $AdS_5\times S^5$ and is given in terms
of a pp-wave metric supported by a self-dual five-form \cite{Blau:2001ne}.
 In the gauge theory side, the limit corresponds to isolate
composite operators with large ${\cal R}$-charge.

Plane wave geometries are exact solutions of string theory (see
\cite{Tseytlin:1995fh} and references therein).
Under certain conditions they lead to exactly solvable string
models where the spectrum and simple correlation functions
can be computed much in the same way as in flat space.
 We are interested in plane waves arising as Penrose limits
of p-brane geometries. The rich isometry group of near horizon
p-brane solutions relates most of the choices for null geodesics
leaving only few independent possibilities. In case one starts with
a pure $AdS\times S$ solution, the two independent possibilities
lead to either a pp-wave or a flat space-time \cite{Blau:2001ne}.

 In \cite{Gimon:2002sf,Fuji:2002vs}, the Penrose limits of non-conformal
Dp brane geometries were
considered. The limit along a generic null geodesic results
again into a pp-wave solution, but now with running
dilaton, masses and RR-fluxes.
Remarkably, as first noticed in \cite{Fuji:2002vs}, the
geometries display a ``critical radius'' which acts as a fixed
point in the radial evolution of a particular class of null geodesics.
  The physics along such ``critical'' geodesics is accurately described by
a solvable string theory with constant masses, dilaton and RR fluxes
and will be the main subject of our study.

  Following the BMN ideas, it is natural to consider
string theories on these solvable pp-wave vacua
as the holographic duals of states with large ${\cal R}$-charge in the
SYM theories leaving on the branes.
 The worldvolume theories leaving on Dp-branes are defined via
dimensional reduction from ${\cal N}=1$ SU(N) SYM in $D=10$ down to $d=p+1$.
 For $p\neq 3$, they are non-conformal gauge theories
and preserve sixteen supercharges.
 Before taking the Penrose limit they are described by warped $AdS$
supergravities.
Various aspects of these correspondences
have been studied in \cite{Itzhaki:1998dd} and
more recently in \cite{Morales:2002ys}.
The Penrose limit, we are considering here, correspond to a very
special limit where ${\cal R}$-charges in the SYM theory
are taken to be large while keeping fixed the working energy scale.
  This is of course the closest analog of the ``conformal''
pp-wave versions of the correspondence.
 The physical meaning of this ``critical energy'' in the SYM side
remains to be fully clarified.

  The aim of this paper is to study the spectrum of
string theory on general Dp-waves.
 We would like to stress that solvable Dp-wave
string models are interesting by itself.
They provide the simplest string vacua with sixteen supercharges and
non-trivial RR fluxes.  In the absence of a better
understanding of string theory on more general RR backgrounds,
pp-wave solvable models constitute the unique tractable examples.
 In the same spirit, string-string dualities
in the presence of RR-fields are fully unexplored and Dp plane
waves provide a simplified setting where a deeper study of this
physics can be addressed.

  In this paper we mainly focus on D1-waves in type IIB/type I theory.
Besides the duality with 2D SYM theories, these models are interesting
because they can be reformulated entirely in terms of NSNS
F1-plane waves
in the S-dual type IIB/heterotic string respectively. The S-dual pairs
are associated to Penrose limits along critical null geodesics
of the corresponding p-brane solutions in the low energy supergravities.
 Type I descendants are constructed by introducing a consistent set
of D9-O9 planes in the type IIB D1-wave. It is important
to notice that D9-O9 systems
are allowed in the D1-wave unlike in the more familiar
D3-wave geometry where they were excluded by
the selection rules derived in \cite{Billo:2002ff,Dabholkar:2002zc,
Bergman:2002hv}.

  The paper is organized as follows: In Section 2 we review and
adapt the results of \cite{Fuji:2002vs} in order to derive
pp-wave solutions as ``critical'' Penrose limits of Dp-brane
near horizon geometries.
Section 3 is devoted to the study of
the spectrum of KK harmonics in ${\cal N}=1,2$ supergravity
on the D1/F1 plane wave. In section 4, we compute one-loop
modular invariant helicity supertraces
counting multiplicities and charges of string excitations
in the Dp-waves.
The formulas encode in a unifying way the resulting
spectrum of S-dual pairs of F1/D1 plane waves in
10D superstrings. The Klein, Annulus and Moebius amplitudes
defining open descendants of D1-waves are also discussed.
 In section 5 we summarize the results and comment on interesting
directions for future research.

\vspace*{.5cm}

\section{Penrose limits of Dp-branes}

Penrose limits of non-conformal
brane solutions have been recently studied in \cite{Gimon:2002sf,Fuji:2002vs}.
For $p\neq 3$ and a generic choice of null geodesics
the limit results into a pp-wave with
$x^+$-dependent dilaton, masses and RR-fluxes.
However, as noticed in \cite{Fuji:2002vs}, to any given
ratio $\ell\equiv {J\over E}$ of the energy
and angular momentum eigenvalues, there exists a critical
radius $r_0$, in the nearby of which, physics is effectively described
by a solvable string theory.
We are interested in Penrose limits along geodesics starting near $r_0$.
In this section we review (and slightly adapt)
the results of \cite{Gimon:2002sf,Fuji:2002vs}
focusing on this particular class of geodesics.
For reasons that will become evident later, we choose to work
in the Einstein frame.

Dp-brane solutions involve, besides the metric, non-trivial profiles
for the dilaton and RR $(p+2)$-form. In the
Einstein frame the solution can be written as :
\bea
ds^2&=
&H^{{p-7\over 8}}\,(-dt^2+d\vec{x}^2)+H^{p+1\over 8}\,(dr^2+r^2\,d\psi^2
+r^2\,{\rm \sin}^2 \psi\,d\Omega_{7-p}^2)\nn\\
F_{p+2} &=& d{\rm vol}({\bf E}^{1,p})\wedge dH^{-1}\nn\\
e^{\phi-\phi_\infty}&=&H^{3-p\over 4}
\label{Dp}
\eea
with $H$ the harmonic function
in the transverse space :
\be
H=1+{Q_p\over r^{7-p}}\approx {Q_p\over r^{7-p}} \quad\quad
\quad\quad Q_p=c_p\, g_{\rm YM}^2\,N (\alpha^\prime)^{5-p}
\ee
Throughout this paper we restrict ourselves to near horizon regions
$r^{7-p}\ll Q_p$, where the one in $H$ can be discarded.

We are interested in the physics along the null geodesics:
\be
r= r(x^+) \quad\quad \psi= -b(x^+) \quad\quad
t= -\ell\, \psi-a(x^+)
\label{geodesics}
\ee
with $r(x^+)$, $a(x^+)$, $b(x^+)$, functions of the
affine parameter $x^+$ defined through the null geodesics
and (J,E)-eigenvalue conditions
\bea
-g_{tt}\dot{t}=1\quad\quad g_{\psi\psi} \dot{\psi}=\ell \quad\quad
g_{tt}\dot{t}^2+g_{\psi\psi}\dot{\psi}^2+g_{rr} \dot{r}^2=0
\label{nullg}
\eea
Dots stand for the derivatives with respect to $x^+$.
The first eq. in (\ref{nullg}) fixes the normalization of
the affine parameter by setting the energy $E=1$, while $\ell=J/E$
parametrizes the angular momentum.
Eqs  (\ref{nullg}) are solved by\footnote{Solutions
can be written in terms of hypergeometric functions, but the
explicit form will not be needed here .}:
\bea
\dot{r} &=& {\omega\over A\, B^2\, r} \quad\quad
\dot{a}= {1\over A^2}-{\ell^2\over B^2 r^2}\quad\quad
\dot{b}=  -{\ell\over B^2 r^2}
\label{sol}
\eea
with $A^2\equiv H^{p-7\over 8}$, $B^2\equiv H^{p+1\over 8}$
and $\omega(x^+)\equiv
\sqrt{B^2 r^2-\ell^2 A^2}$.
All functions in (\ref{sol}) carry an implicit dependence on $x^+$
through $r(x^+)$.

In order to define the Penrose limit it is convenient to
introduce the following change of coordinates:
\bea
\vec{x}&\rightarrow & \Omega {\bf x}/A \quad\quad
\vec{y}\rightarrow \Omega {\bf y}/(Br)\quad\quad
r\rightarrow r(x^+)\quad\quad
\psi \rightarrow  -b+ \Omega\, z\,/ \omega\nn\\
t &\rightarrow & -a-\ell \, \psi+\Omega^2\,\left[ x^- +
{\partial_+ \omega\over 2 \omega}\, z^2 +
{\partial_+ A\over 2 A}\,{\bf x}^2+
 {\partial_+ (B r \sin{b})\over 2 B  r \sin{b}}\,{\bf y}^2
\right]
\label{change}
\eea
which reduces at $\Omega\rightarrow 0$ to the
null geodesics (\ref{geodesics}).
 We parametrize by $\vec{y}$ the coordinates on
the unitary sphere $S^{7-p}$. In the limit $\Omega\rightarrow 0$
they describe a point in ${\bf R}^{7-p}$ with coordinates ${\bf y}$.
 By plugging (\ref{change}) into (\ref{Dp}) and keeping only
the leading terms in $\Omega\rightarrow 0$ one finds:
\bea
\Omega^{-2}\,ds^2 &=&
2\,dx^+\,dx^- -(m_x^2 {\bf x}^2+m_z^2 z^2+m_y^2 {\bf y}^2)\,dx^{+2}
+dz^2+d{\bf x}^2+d{\bf y}^2\nn\\
\Omega^{-(p+1)} F_{p+2} &=& 2\hat{f} \,
dx^+\wedge d{\rm vol}({\bf E}^p)\wedge dz
\label{vacuum}
\eea
with
$e^{\phi-\phi_\infty}=\left({Q_p\over r^{7-p}}\right)^{3-p\over 4}$ and
\bea
m_{x}^2 &=& m_{z}^2= {(7-p)^2\over 256} Q_p^{-{p+1\over 4}}\,
r^{-{(p-3)^2\over 4}}\,\left[\ell^2\,(31-2p-p^2)+
(p-3)^2\, Q_p\,r^{p-5}\right]\nn\\
m_{y}^2 &=&  {(7-p)\over 256} Q_p^{-{p+1\over 4}}\,
r^{-{(p-3)^2\over 4}}\,\left[\ell^2\,(25+19p-5p^2+p^3)+
(p-3)^2(7-p) Q_p\,r^{p-5}\right]\nn\\
2\hat{f} &=&(7-p)\ell\, Q^{-11+4p-p^2\over 16} r^{(3-p)^2(5-p)\over 16}
\label{masses}
\eea
For $p=3$ one recovers the familiar pp-wave solution with masses
$m_x^2=m_y^2=m_z^2=\ell^2/Q_p$. On the other hand,
for $p\neq 3$, the
solution (\ref{vacuum}) involves non-trivial $x^+$-dependent dilaton,
masses and RR-fluxes.
 A simple inspection of (\ref{sol}) reveals however that
for $p\neq 5$ the radial evolution equation has a fixed point at the
critical radius \cite{Fuji:2002vs}\footnote{The case $p=5$ should be
treated independently and it will be omitted here (see
\cite{Fuji:2002vs,Oz:2002ku} for discussions on this case)}
\be
r_0^{5-p}\approx {Q_p\over \ell^2}
\label{r0}
\ee
where $\dot{r}=0$.
Geodesics starting near this critical radius
keeps close to it along $x^+$-evolution leading to a pp-wave
solution with constant masses, dilaton and RR-fluxes \cite{Fuji:2002vs}.
Evaluating  (\ref{masses}) at $r_0$ we are finally left with
\bea
m_{x}^2&=& m_z^2=\ft{1}{32}\,(7-p)^2\,(5-p)\,m^2\nn\\
m_{y}^2&=&\ft{1}{32}\,(7-p)\,(11-4p+p^2)\,m^2\nn\\
e^{a_p \phi_0 \over 2}\,2 \hat{f}&=&(7-p)\,m \label{msf}
\eea
with $m^2=\ell^{29-10p+p^2\over 2(5-p)} Q_p^{7-p\over 2(p-5)}$ and
constant dilaton
$e^{a_p\phi}=e^{a_p\phi_0}= (Q_p \ell^{p-7})^{(3-p)^2\over 4(5-p)}$.

 One can directly verify that (\ref{vacuum}) is indeed a solution of
the ten dimensional Einstein equations. The only non-trivial
component of the Einstein-Hilbert system:
\be
R_{++}={ e^{a_p\mathring{\phi}}\over 2 n!}
\left(n\, F_{+A_2....A_n}\, F_+^{A_2....A_n}-
{n-1\over 8} g_{++} F^2  \right)
\ee
requires
\be
\sum_i\, m_i^2=(p+1)\,m_x^2+(7-p)\,m_{y}^2=
2\,e^{a_p \phi_0} \hat{f}^2
\ee
which is verified by (\ref{msf})\footnote{
In the following we will always reabsorve
the $e^{a_p \phi_0\over 2}$ factor in the
definition of $\hat{f}$ setting $e^{a_p \phi_0}=1$.}
Notice that the expressions (\ref{msf}) for the pp-wave masses differ
substantially from those obtained in \cite{Fuji:2002vs} using
the string frame metric.
This is not surprising since the two frames are related by a
non-trivial $x^+$-dependent redefinition. We should stress
that our choice here of working in the
Einstein frame is not ``ad hoc''. Of course solutions coming
from the Penrose limit (at the critical radius) using the string
metric also fulfill the Einstein equations
but due to the non-canonical kinetic term $e^{-2\phi}\, {\cal R}$,
a bit more care should be taken in discarding $\partial_+^2  \phi$
terms.

For $p\neq 3$ the solutions (\ref{vacuum}) preserve half of the
supersymmetries. Killing spinors equations can be solved
in terms of spinor components $\epsilon$ satisfying
$\Gamma^+ \epsilon=0$ \cite{Blau:2001ne}. This matches
the number of supersymmetries of the SYM theories living on the Dp-branes.

\section{KK supergravity harmonics}

In this section we compute the bosonic spectrum of type II,I
supergravities in the F1/D1-wave geometry.
We follow closely \cite{Metsaev:2002re} where similar results were
derived in the D3-wave case. The KK spectra found here
should correspond to contractions of those in \cite{Morales:2002ys},
where the type IIB/I linearized equations were solved
in the warped $AdS_3\times S^7$ near horizon geometries.
 It would be nice to have a more concrete realization of this idea.

 For definiteness we focus on the F1-pp wave case. $SL(2,\bbZ)$
invariance of type IIB supergravity makes this
vacuum  indistinguishable from that of the D1-pp wave and therefore the
two spectra are identical. States will be
classified according to the $SO(6)$ R-symmetry group acting on
the transverse ${\bf R^6}$.
This group is identified with the subgroup of the $SO(8)$
${\cal R}$-symmetry group in the two-dimensional gauge theory
that survives the large R-charge BMN limit.
In addition states will come
in groups of $256$ states at each level of the KK tower.
This was also the case in the warped geometry
\cite{Morales:2002ys} before taking the Penrose limit.

The spectrum of KK harmonics is determined by solving the
linearized type IIB equations of motion in the pp-wave
geometry. To linear order in the fluctuations around
(\ref{vacuum}), type IIB supergravity fields can be written as
(we borrow the notations of \cite{Schwarz:qr,Morales:2002ys}):
\bea
g_{MN} &=&\mathring{g}_{MN}+h_{MN}\nn\\
P_M &=&i\, {\partial_M\, \tau \over 2 \tau_2}=
{1\over 2}\,\partial_M\, \Phi\nn\\
q_M &=&-{\partial_M \tau_1\over 2\tau_2}=-{e^{\phi_0}\over 2}\partial_M C_0\nn\\
G_{MNP}&=&\mathring{G}_{MNP}+{\cal G}_{MNP}= {\varphi\over 2}\, \mathring{G}_{MNP}+3\partial_{[M}\,a_{NP]}\nn\\
f_{MNPQR}&=&5\,\partial_{[M}\,a_{NPQR]}+
\ft54\left(a^{1}_{[MN}\partial_{P}\,a^2_{QR]}-
a^{2}_{[MN} \mathring{G}_{PQR]}\right)
\label{fluctuations}
\eea
where we have introduced the complex notation
\bea
\Phi&=&\Phi_1+i\Phi_2=\varphi+i e^{\phi_0} \chi \nn\\
a_{MN}&=&a^1_{MN}+i a^2_{MN}=
e^{\phi_0\over 2 }\, a_{MN}^{NS}+i e^{-{\phi_0\over 2} }\, a_{MN}^{R}
\eea
for fluctuations of the dilaton-axion and two-form systems respectively.
In addition we will impose the light cone gauge conditions:
\be
h_{-M}=a_{-M}=a_{-MNP}=0
\ee
to get rid of gauge degrees of freedom in the metric, two and four form.
We will start by collecting the equations of motion for the various
fields, which upon diagonalization reduce to a set of
second order differential equations of the kind
\be
\left(\square + i\lambda_s \, \partial^+\right)\, \Phi_s=0
\label{diffeq}
\ee
with
\be
\square\equiv g^{MN}\,\partial_M\partial_N=
2\partial_+\partial_- +\partial_I^2
-m_I^2\, y^{I\,2}\, \partial_-^2    \;.
\ee
The index  $s=1,.., 256$ running over the 10D type IIB supergravity modes.
We collect in $y^I=\{ {\bf x}, z, {\bf y} \}$, $I=1,...8$, the
transverse to the light-cone coordinates and denote by $m_I$ their
associated masses.
Identifying as in \cite{Metsaev:2002re} our
Hamiltonian by $H=-i\partial^-$, eq.
(\ref{diffeq}) reduces to a system of eight harmonic oscillators
with masses $m_I$ and energies
\be
E_s=\lambda_s-a+\sum_{I=1}^8 m_I n_I
\ee
with $n_I$ the oscillator occupation number and $a$ the zero point energy.
$\lambda_s$
determine the shift in energy of the field $\Phi_s$
respect to other components at the same level in the KK-tower.
The main task is to determine the spectrum of $\lambda$'s by
diagonalizing the linearized supergravity equations.
Readers that are not interested in the details of this
derivation can skip the next subsections directly
to table 1, where the final results are summarized. \\

\subsection{Type IIB supergravity}

We start by considering the linearized equations of motion
for type IIB supergravity around the F1-wave.

{\bf Scalar Equation:}
\\
The dilaton-axion equation of motion in the plane wave background
reads:
\be
\mathring{D}{}^M\, P_M=\ft14  \mathring{G}{}^{-IJ}{\cal G}_{-IJ}
\label{eq0}
\ee
In the light-cone gauge one has
 ${\cal G}_{-IJ}=\partial^+\, a_{IJ}$ and (\ref{eq0}) can be
rewritten as:
\be
\square\, \Phi-\ft12 c^{IJ}\,\partial^+ a_{IJ}=0
\label{s0}
\ee
with $c_{IJ}=\mathring{G}_{+IJ}$ the three-form flux.
\\\\
{\bf Three-form Equation:}
\\
The linearized equations for the complex three-form equations read:
\be
\delta D^P \mathring{G}_{MNP}+\mathring{D}{}^P\, {\cal G}_{MNP}=
-P^P\,\mathring{G}{}^*_{MNP}
-\ft23 i\, f_{MNPQR}\,\mathring{G}{}^{MNP}
\label{e3}
\ee
As in \cite{Metsaev:2002re} it is enough to
solve for the components $(-I)$ and $(IJ)$.
The $(-I)$-component eqs:
\be
 \mathring{D}^P\, {\cal G}_{-IP}=0\quad\quad \Rightarrow ~
\partial^P\,a_{PI}=0
\ee
can be used to rewrite
\bea
\mathring{D}{}^P\, {\cal G}_{IJP}&=&\square\, a_{IJ}
+\ft12 c_{IJ}\partial^+\varphi\nn\\
\delta D^P \mathring{G}_{IJP} &=&\ft{i}{2}\,
c_{IJ} \partial^+ \Phi_2
+2 \partial^+ h_{K[I}\,c_{J]K}
\eea
By plugging this in (\ref{e3}) and after some simple algebra one finds
\be
\square\, a_{IJ}+\partial^+ (c_{IJ}\, \Phi+2 \, h_{K[I}\,c_{J]K}+2i\, c^{KL}\,
\hat{a}_{IJKL})=0
\label{s1}
\ee
with $\hat{a}_{I_1..I_4}\equiv a_{I_1..I_4}+a^{1}_{[I_1I_2}\,a^2_{I_3I_4]}$.
\\\\
{\bf Metric Equations:}
\\
The linearized graviton equations around the pp-wave geometry can
be writen as
\be
r_{MN}=\ft18\left(  \mathring{G}_{M}{}^{PQ}{\cal G}^*_{NPQ}+
\mathring{G}_{N}{}^{PQ}{\cal G}^*_{MPQ}-\ft16\, \mathring{g}_{MN}
\mathring{G}{}^{PQR}{\cal G}^*_{PQR}+{\rm h.c.}\right)
\label{eqs2}
\ee
As in \cite{Metsaev:2002re} the components $r_{--}=r_{-I}=0$ gives
$h_{II}=D^m\,h_{mI}=0$. This allows us to rewrite $r_{IJ}=-\ft12
\square \, h_{IJ}$ and (\ref{eqs2}) reads:
\be
\square\, h_{IJ}+\ft12 \partial^+ (c_{IK}\,a_{JK}^*+c_{JK}\,a_{IK}^*
- \ft14\,\delta_{IJ} c_{KL}\, a^*_{KL}+{\rm h.c.})=0
\label{s2}
\ee
{\bf Five-form equations}\\
Self-duality of the $(-I_1...I_4)$ components gives
\be
\hat{a}_{I_1..I_4}={1\over 4!} \epsilon_{I_1..I_4J_1..J_4}\, \hat{a}_{J_1..J_4}
\ee
 Using the self-duality of the five-form and the definition
(\ref{fluctuations}) one writes
\be
\mathring{D}{}^P\, f_{PM_1..M_4}={1\over 2\, 3!^2}\, \epsilon_{M_1....M_{10}}\,
F_{1}^{M_5M_6M_7}F_{2}^{M8M_9M_{10}}
\label{e5}
\ee
In particular for the $(-I_1..I_3),(I_1,...I_4)$ components one has:
\bea
\mathring{D}{}^P\, f_{P-I_1..I_3}&=&0 \Rightarrow \partial^P\,
\hat{a}_{P I_1..I_3}=0\nn\\
\mathring{D}{}^P\, f_{PI_1..I_4}&=&{1\over 8}\, \epsilon_{I_1..I_4J_1..J_4}\,
F_{1}^{-J_1J_2}F_{2}^{+J_3J_4}
\eea
Plugging in (\ref{e5}) one finally finds:
\be
\square\, \hat{a}_{I_1..I_4}-\ft12\partial^+ \left[ c_{[I_1I_2}a_{2I_3I_4]}
+{1\over 4!}\epsilon_{I_1...J_4}c^{[J_1J_2}a_2^{J_3J_4]}\right]=0
\label{s5}
\ee
{\bf Spectrum of KK States:}\\

We now specify to a NSNS two-form flux $c_{12}=2 \hat{f}$
along the $(12)$ plane.
We split capital indices $I$ into $\alpha,\beta=1,2$, $i,j=1,...6$,
according to whether they are parallel or perpendicular to the flux.
Collecting equations (\ref{s0}),(\ref{s1})
(\ref{s2}) and (\ref{s5}), we find upon diagonalization a system
of differential equations of the kind (\ref{diffeq}) with
$\lambda$-eigenvalues displayed in table \ref{tableII}.

\begin{table}[h]
\centering
 \begin{tabular}{llll}
 Fields & $SO(6)$ & d.o.f. & $\lambda$  \cr
 $h$ & ${\bf 1}$ & 1 & -$4 \hat{f}$  \cr
 $a_{\alpha i}^{NS},\chi, a_{ij}^{R}$
& ${\bf 1}+2\times {\bf 6}+{\bf 15}$ &28 & -$2 \hat{f}$ \cr
$h_{12}, \bar{h}_\parallel, h_\parallel, h_{ij}, a_{ij}^{NS}, a_{\alpha i}^{R},a_{\alpha i_1..i_3}$
& $3\times {\bf 1}+2\times {\bf 6}+{\bf 15}+
2\times {\bf 20}$&70 & 0 \cr
 $\bar{a}_{\alpha i}^{NS},\bar{\chi}, \bar{a}_{ij}^{RR}$
& ${\bf 1}+2\times {\bf 6}+{\bf 15}$ &28 & $2  \hat{f}$ \cr
 $\bar{h}$ & ${\bf 1}$ & 1 & $4   \hat{f}$
\end{tabular}
\caption{KK bosonic spectrum of Type II supergravity.}
\label{tableII}
\end{table}

We denote by $a_{\alpha i},\chi,h, h_\parallel$ some
complex linear combinations of the axion/dilaton, metric and NSNS/RR forms.
We display only the quantum numbers of the $128$ bosonic states
at the level 0 in the KK tower, i.e. in the bosonic zero mode ground state.
States in the higher KK floors are still organize in groups of $256$ states
($128$ bosons and $128$ fermions). Energies are shifted
by $n_I m_I$. $SO(6)$-representations of states in this higher KK levels
can be found by tensoring the representations in the
table with those of the harmonic oscillator zero modes $a_0^I$.

\subsection{Type I supergravity}

The analysis above can be applied ``mutatis mutandis'' to the study of
D1/F1-waves in type I/heterotic theory. The spectrum of KK states
in the heterotic F1-wave
can be read from that of type IIB by simply projecting out RR states
and adding the contribution of $n_V$ ${\cal N}=1$ vector multiplets.
Similarly states in the
type I D1-wave are related to those in the heterotic via S-duality.

The result of the $\Omega$-projection is displayed in table \ref{tableI}.

\begin{table}[h]
\centering
\begin{tabular}{llll}
 Fields & $SO(6)$ & d.o.f. & $\lambda$  \cr
 $h$ & ${\bf 1}$ & 1 & - $4\hat{f}$ \cr
 $a_{\alpha i}$
& $2\times {\bf 6}$ &12 & -$2\hat{f}$ \cr
$h_{12}, \bar{h}_\parallel, h_\parallel, h_{ij}, a_{ij}$
& $3\times {\bf 1}+{\bf 15}+{\bf 20}$&38 & 0 \cr
 $\bar{a}_{\alpha i}$
& $2\times {\bf 6}$ &12 & $2\hat{f}$ \cr
 $\bar{h}$ & ${\bf 1}$ & 1 & $4\hat{f}$
\end{tabular}
\caption{Type I Supergravity multiplet.}
\label{tableI}
\end{table}

  In addition we have $n_V$ ${\cal N}=1$ vector multiplets
with vector equations
\be D_M\, F^{MN}-\ft12
\mathring{G}_{NPQ}\, F^{PQ}=0 \;,
\label{typeIeq}
\ee
and $F_{MN}\equiv
2 e^{\mathring{\phi}\over 2}
\partial_{[M}\,A_{N]}$. The equation for the $(-)$-component gives
\be
 \mathring{D}^P\, F_{P-}=0\quad\quad \Rightarrow ~
\partial^P\,a_{P}=0
\ee
 while for $I$ components we find
\be
\square\, a_{I}-c_{IJ} \,\partial^+ a_J=0
\label{g1}
\ee
Upon diagonalization, (\ref{g1}) reduces to our familiar
differential equation with $\lambda$-eigenvalues given in
table \ref{tableGI} \\
\begin{table}[h]
\centering
 \begin{tabular}{llll}
 Fields & $SO(6)$ & d.o.f. &  $\lambda$  \cr
 $a$ & ${\bf 1}$ & 1 & -$2\hat{f}$   \cr
 $a_{i}$
& ${\bf 6}$ &6 & 0 \cr
 $\bar{a}$ & ${\bf 1}$ & 1 &$2\hat{f}$
\end{tabular}
\caption{Type I Gauge multiplet.}
\label{tableGI}
\end{table}
  As before, states in the higher levels of the KK tower
can be found by acting with the oscillator bosonic zero modes
$a_0^I$ and carry energies
shifted by $n_I m_I$. Notice that now at each KK level,
states are organized in groups of $128=64_B+64_F$ uncharged components and
adjoint multiplets containing
each $16=8_B+8_F$ physical degrees of freedom.

\section{String theory on Dp-plane waves}

In this section we compute the helicity supertrace of string theory
on Dp-waves (\ref{vacuum}) with
masses $m_I$ and RR/NSNS fluxes $\hat{f}$.
We follow closely \cite{Russo:2002rq}, where similar techniques were
applied in the six-dimensional context (see also
\cite{Kiritsis:2002kz} for related discussions).
 For definiteness we focus on type IIB.
 The general coupling of strings moving in a plane wave with NSNS/RR
constant fluxes can be summarized in the following
worldsheet Lagrangian:
\be
{\cal L}= {\cal L}_0+{\cal L}_R+{\cal L}_{NS}
\ee
where
\bea
{\cal L}_0&=& {1\over 2}\left(\partial_+ Z^i \partial_- \bar{Z}^i-
\mu_i^2 Z^i \bar{Z}^i+{\rm h.c.}\right)+S^a\partial_+ S^a+
 \tilde{S}^a\partial_- \tilde{S}^a\nn\\
{\cal L}_R&=& -2i f\, S^a \Pi_{ab} \tilde{S}^b\nn\\
{\cal L}_{NS}&=& {f_i\over 2}\left(i\bar{Z}^i \partial_+ Z^i-
\bar{Z}^i \partial_- Z^i+ {\rm h.c.}\right)- {if_i\over 2}\left(
S \gamma^{i\bar{i}}S-
\tilde{S}^a \gamma^{i\bar{i}}_{ab} \tilde{S}^b\right)
\eea
We use the lightcone gauge $x^+=\alpha^\prime p^+ \sigma_0$
and $\partial_{\pm}=\partial_0\pm \partial_1$.
Worldsheet masses are related to the spacetime variables via
$\mu_i=2 p^+\alpha^\prime m_i$, $f=
2 p^+\alpha^\prime \hat{f}$, $f_i=
2 p^+\alpha^\prime H_{2i 2i-1 +}$.

The Lagrangian densities
${\cal L}_f$ (${\cal L}_v$) represent the coupling to RR(NSNS)
fluxes while ${\cal L}_0$ describe a free string
moving on a pp-wave. We have collected the eight real bosons into four
complex ones and denote their masses by $\mu_i$. Sum over
repeat indices $i,j=1,...4$ is implicitly understood.
The RR fluxes are described by $f$ and couple to fermions
through the mass term $\Pi=\gamma_1....\gamma_{p+1}$.

NSNS fluxes are specified by $f_i$'s.
We have chosen $B_{+i}=\ft{i}{2} \hat{f}_i \bar{z}^i$,
$B_{+\bar{i}}=-\ft{i}{2} \hat{f}_i z^i$ such that
$H_{i\bar{i}+}=\hat{f}_i$.
Notice that the coupling to fermions of NSNS
fields are substantially different from that of RR-fields.
Indeed the
whole ${\cal L}_{NS}$-Lagrangian
can be reabsorved by the worldsheet redefinition \cite{Russo:2002rq}
\be
Z^i \rightarrow e^{i f_i \sigma_1} Z^i \quad\quad
S \rightarrow e^{{i\over 2} f_i\sigma_1
\gamma^{i\bar{i}} } S    \quad\quad  \tilde{S}\rightarrow
e^{{i\over 2}\sigma_1 f_i \gamma^{i\bar{i}} } \tilde{S}
\label{twist}
\ee
and $\mu_i^2 \rightarrow \mu_i^2-f_i^2$.
The new fields satisfy twisted boundary
conditions specified by $v_i\equiv a_i \tau+b_i=i f_i \tau_2$.

  In the following we will restrict ourselves to RR-waves, but
we will keep track of ${\cal R}$-charges by computing the one-loop
partition functions in a given spin structure specified by $v_i=\bar{v}_i$.
 For future references we stress that the resulting helicity traces
can be used, according to our discussion above, to extract
one-loop correlation functions involving an arbitrary number of insertions
of the NSNS B-fields on the RR plane wave.

 We start by considering oriented closed string states.

\subsection{The torus amplitude}

One-loop partition functions for type IIB on D3 pp-waves
have been computed in \cite{Russo:2002rq,Takayanagi:2002pi,
Sugawara:2002rs}.
In this section we compute helicity supertraces for
string theories on Dp-plane waves.
 Helicity supertraces weight states according to their R-symmetry charges
and allow us to perform a more refined comparison
of multiplicities and charges with gauge/supergravity descriptions.
 In addition, as we have discussed earlier in this section,
they provide a unifying description of D1/F1 plane wave
string vacua.

 An important new feature of string theory on general Dp-waves is that
bosons and fermions carry in general different worldsheet masses
$\mu_i, f$.
 The Lorentz $SO(8)$ symmetry is broken
by pp-masses and RR-fluxes to its $SO(p+1)\times SO(7-p)$ subgroup.
The helicity supertrace measures string multiplicities and
charges under the $SO(2)^4$ Cartan subgroup
of $SO(p+1)\times SO(7-p)$. More precisely, we define
\footnote{We borrow the notations
of \cite{Sugawara:2002rs}.}
\be
Z(v_i|\tau,\bar{\tau})={\rm Tr}\, (-)^F\, e^{-2\pi \tau_2 H+
2\pi i\,\tau_1\,P +2\,\pi\,v^i\, J_i}
\label{trace}
\ee
where
\bea
H &=& \sum_{n\in \bbZ}\left(\omega^i_n \, N_{Bn}^i+\omega^f_n\, N_{Fn}\right)
-a  \quad\quad a=2 \,\sum_{i=1}^4\, (\Delta_f-\Delta_{\mu_i})
\nn \\
P&=& \sum_{n\in \bbZ} n\, N_n\quad\quad
\omega^i_n=\sqrt{\mu_i^2+n^2},~\omega^f_n=\sqrt{f^2+n^2}\nn\\
v_i\, J^i&=& \pm \sum_{n\in \bbZ}\, \left(v_i\, N_{Bn}^i+
\tilde{v}_i \, N_{Fn}^i\right).
\eea
We have built with ${\bf x},z,{\bf y}$ four complex
bosons and assigned them masses $\vec{\mu}$ according to (\ref{msf}).
In addition we have introduced the
quantities $\tilde{v}_i$ defined through:
\bea
\tilde{v}_1 &=&\ft12(v_1+v_2+v_3+v_4)\nn\\
\tilde{v}_2 &=&\ft12(v_1+v_2-v_3-v_4)\nn\\
\tilde{v}_3 &=&\ft12(v_1-v_2-v_3+v_4)\nn\\
\tilde{v}_4 &=&\ft12(v_1-v_2+v_3-v_4)
\eea
which characterize the $SO(8)$ spinor representation
under which worldsheet fermions transform.

The presence of the extra v-term twists the boundary conditions of
worldsheet bosons and fermions by
$v_i$ and $\tilde{v}_i$ respectively. The twist is now along $\sigma_0$
in contrast with that in (\ref{twist}).
The result follows with minor modifications from that
in \cite{Takayanagi:2002pi,Sugawara:2002rs}:
\be
Z(\vec{v}|\tau,\bar{\tau})=
\prod_{i=1}^4{\Theta_{0,\tilde{v}_i}(f|\tau,\bar{\tau})\over
\Theta_{0,v_i}(\mu_i|\tau,\bar{\tau})}\,
\label{helicity}
\ee
An alternative derivation of this formula using the Lagrangian formalism
is presented in the Appendix.
At $v_i=0$, (\ref{helicity}) reduces to the string partition function.
Notice that unlike in the more familiar
D3-brane case, bosonic and fermionic determinants
do not cancel against each other even in the $\vec{v}=0$ case.

Focusing on the zero mode part of (\ref{helicity}):
\bea
{\cal Z}_0(\vec{v},\tau)&=&\prod_{i=1}^4\, e^{2 \pi \tau_2(f-{\mu_i})}
\,\left|{1-q^f e^{2\pi i \tilde{v}_i}\over
1-q^{\mu_i} e^{2\pi i v_i}}\right|^2
\label{z0v}
\eea
with $q=e^{2\pi \tau}$,  one
easily recognizes the spectrum of Kaluza Klein supergravity states
derived in the harmonic analysis above.
The expansion of the numerator in powers of $q$ comprises $256$
states out of which $128$ are bosonic and $128$ are fermionic states.
More precisely besides a singlet state at level 0,
we find  $8$ states in the ${\bf 8}_s$ spinor of $SO(8)$
at level one (term $q^{f}$),
 $28={\bf 8}_s\wedge {\bf 8}_s$ states at level two (order $q^{2f}$)
and so on.
Setting $p=1$, decomposing $SO(8)$ quantum numbers under $SO(6)$ and keeping
only bosonic components, the spectrum in table 1 is reproduced.
Different levels in the Kaluza-Klein tower
are reached by expanding the denominator in (\ref{z0v}).
This corresponds to act with bosonic zero mode creation operators.
Theses terms explicitly break the $SO(8)$ symmetry down to $SO(6)$.
 Degeneracies of these states are in agreement with that of the
set of harmonic oscillators with masses $\mu_i$ found in the
supergravity description.

It is also not hard to associate to each state in (\ref{helicity})
a BMN operator of large ${\cal R}$-charge in the
dual $p+1$-dimensional gauge theory. The dictionary
follows straight from that in \cite{Berenstein:2002jq} by simply
reorganizing the spectrum of operators respect to $SO(p+1)\times SO(7-p)$
rather than $SO(4)\times SO(4)$.
The string vacuum
is again identified with the ${1\over \sqrt{J} N^J}\,{\rm Tr} Z^J$
operator. The $256$ operators associated with states
at a given ``level'' in the KK tower are related by actions of
the eight fermion gaugino components $\chi_{J={1\over 2}}$
in the $d=p+1$-dimensional theory.
The main difference with the $p=3$ case, is that insertions
in the traces of
operators $D_{\alpha} Z$ and $\phi^i$ ($\alpha=1,..p+1$,
$i=1,..7-p$) carry different energies
due to the difference in the longitudinal and transverse masses.
 This is related to the fact that the radii of the original
$AdS_{p+2}$ and $S^{8-p}$ are different.
 It would be nice to explain this rich structure of the energy spectrum
in the gauge theory side.

Before leaving this section it is
worth to spend some words in the flat space limit of our formulas.
For this purpose it is useful to come back to the derivation of the
helicity trace presented in the Appendix.
 As it has been recognized in various places in the literature
(see \cite{Bergman:2002hv,Hammou:2002bf} for example),
the massless limit is far from smooth
due to non-standard bosonic zero mode contributions implicitly
taken into account in (\ref{helicity}).
 More precisely, for $\mu=0$ the computation of the
bosonic determinant (\ref{determinant}) in the Appendix, should be
replaced by that of ${\rm det}^\prime \, \square_v$ where the
$n_1=n_2=0$ constant mode is omitted. The integration over this
constant mode $x_0$ gives the regularized volume of the target
space. A sensible comparison with flat space results is better
performed in terms of the primed partition function \be
Z^\prime(\vec{v}|\tau,\bar{\tau})\equiv
Z(\vec{v}|\tau,\bar{\tau})\, {\cal V}_0(\tau_2)=
Z(\vec{v}|\tau,\bar{\tau})\,\tau_2^{-4}\,\prod_{i=1}^4 |v_i+i
\mu_i \tau_2|^2 \label{prime} \ee where contributions coming from
the constant mode are omitted and collected in ${\cal
V}_0^{-1}(\tau_2)$. Notice that both $Z$ and $Z^\prime$ are
modular invariant since ${\cal V}_0(\tau_2)$ is invariant by
itself under the modular transformation (\ref{modular}).
 In particular for $\vec{v}=0$ the two definitions differ
by powers of the modular invariant combination ${\mu_i^2 \tau_2}$
\footnote{The origin of this normalization factor can be traced in
the ratio between the partition function of a 2d harmonic
oscillator ${1\over \mu^2 \tau_2^2}=\int d{\bf x}_0 d{\bf p}\,
e^{-\pi\tau_2\,( {\bf p}^2+\mu^2 {\bf x}_0^2)}$ and its free
particle limit ${V\over \tau_2}=\int d{\bf x}_0 d{\bf p} \,
e^{-\pi\tau_2 \, {\bf p}^2}$.}.
  In terms of (\ref{prime}) the torus amplitude can be written as
\be
{\cal T}=\int {d^2\tau\over \tau_2^2}\, dp^+\,dp^- e^{-2\pi i \alpha^\prime
p^+\,p^-\,\tau_2}\, {\cal V}_0(\tau_2) \,
Z^\prime(\vec{v}|\tau,\bar{\tau})
\ee
The flat space limit is recovered by sending
$\mu\rightarrow 0$:
\be
\lim_{\mu\to 0}\, Z^\prime (\vec{v}|\tau,\bar{\tau})=
{1\over \tau_2^4}\prod_{i=1}^4
\,\left|v_i\,{ \vartheta_1(\tilde{v}_i) \over \vartheta_1(v_i)}\right|^2
\ee
and replacing ${\cal V}_0(\tau_2)$ by the
regularized volume $V_0$ of the target space. According to
\cite{Sinha:2002di,Hammou:2002bf}
integration over $p^\pm$ set $p^+$ (and therefore $\mu,f$) to
zero and therefore the vacuum energy vanishes at $\vec{v}=0$ as expected.

\subsection{Unoriented and open strings}

In this section we consider open descendants of type IIB
superstrings on D1-waves. Open descendants of
type IIB on D3-waves were first studied in \cite{Berenstein:2002zw}.
As in that case the open string descendant
here will be constructed by adding Dp-Op worldsheet boundaries to the
D1-wave type IIB vacuum.
 We follow standard open string descendant techniques (for a review and
a complete list of references see \cite{Angelantonj:2002ct}.)
 Quantization of open strings moving on RR-plane waves have been
studied in various contexts \cite{Billo:2002ff,
Dabholkar:2002zc,Bergman:2002hv,open}.
The selection rules for the allowed Dp-Op systems
in a D1-wave geometry follow from those been derived in
\cite{Billo:2002ff,Dabholkar:2002zc,Bergman:2002hv}.
 It is convenient to introduce the Dp-mass operator
$M \equiv \Pi\,\Pi_p $  , where $\Pi=\gamma^{12}$ and
$\Pi_p=\gamma^{12..p-1}$ are the chirality operators along the
D1-flux plane and Dp/Op plane respectively. A consistent Dp/Op
system is defined roughly by requiring that ${\rm Tr}\, M=0$ and
$M^2=-1$ \footnote{These two conditions ensure that the effective
zero mode Hamiltonian $H\sim S_0^a M_{ab} S_0^b$ reproduces the
right fermionic contributions in the closed string channel
\cite{Bergman:2002hv}}. This implies that the difference between
the number of NN and DD directions is $2~mod~ 4$.
  These conditions are clearly satisfied by Dp-Op systems
with p=5,9 oriented along $x^{\pm}$ and parallel to the D1-plane
flux.
 We will focus on these two cases, although our formulas are derived
in a rather more general perspective.
These systems are of particular interest since they should presumably
provide S-dual string duals of heterotic string on F1-plane waves.

 Open string amplitudes and tadpole conditions
for self-dual pp-wave vacua have been
worked out recently in \cite{Sinha:2002di,Hammou:2002bf}.
 The extension of these results to our case is
straightforward.
  Apart from the zero mode part
the Klein and open string
amplitudes are roughly the square root of (\ref{helicity}):
\bea
{\cal K} &=& \int {d^2\tau\over \tau_2^2}\, dp^+\,dp^- e^{-2\pi \alpha^\prime
p^+p^-\tau_2}\, {Z^{1\over 2}(2i\tau_2)
 \over Z_0^{1\over 2}(2i\tau_2)}\,{\cal K}_0\nn\\
{\cal A} &=& N^2 \int {d^2\tau\over \tau_2^2}\, dp^+\,dp^-\,
e^{-2\pi  \alpha^\prime
p^+p^-\tau_2}\, {Z^{1\over 2}({i\tau_2\over 2})
 \over Z_0^{1\over 2}({i\tau_2\over 2})}\, {\cal A}_0\nn\\
{\cal M} &=& \pm N\int {d^2\tau\over \tau_2^2}\, dp^+\,dp^-
e^{-2\pi  \alpha^\prime
p^+p^-\tau_2}\, {Z^{1\over 2}(
{i\tau_2\over 2}+{1\over 2})
 \over Z_0^{1\over 2}(\ft{i\tau_2}{2}+ \ft12)}\,{\cal M}_0
\label{direct}
\eea
In (\ref{direct}) and throughout this section we have settled $\vec{v}=0$.
By $N$ we denote the number of branes, while $\pm$ denote the possible
orthogonal or symplectic choice of Moebius projections in the
open string sector. Finally we have denoted by ${\cal Z}_0$, ${\cal K}_0$,
 ${\cal A}_0$ ${\cal M}_0$ the zero mode contributions to the
torus, Klein, Annulus and Moebius strip amplitudes respectively:
\bea
{\cal Z}_0(i\tau_2)&=&\prod_{i=1}^4\, q^{(\mu_i-f)}
\,{(1-q^f)^2\over
(1-q^{\mu_i})^2} \nn\\
{\cal K}_0 &=&\prod_{i=1}^4\, q^{(\mu_i-f)}
\,{(1-q^{2f})\over
(1+\epsilon_i q^{\mu_i})^2} \nn\\
{\cal A}_0={\cal M}_0 &=&\prod_{i=1}^4\,
q^{\ft14(\epsilon_i+1)\mu_i-\ft14 f} \,{(1-q^{f\over 2})\over
(1-q^{\mu_i\over 2})^{\epsilon_i+1}} \eea  Here $q=e^{-2\pi
\tau_2}$ and $\epsilon_i=+,-$ for NN and DD directions
respectively, i.e. \bea {\rm D9/O9} && \quad\quad
\epsilon_1=\epsilon_2=\epsilon_3=\epsilon_4=+1\nn\\
{\rm D5/O5} && \quad\quad \epsilon_1=\epsilon_2=+1\quad
 \epsilon_3=\epsilon_4=-1
\eea The $\epsilon_i$ in the Klein bottle amplitude reflects the
minus eigenvalue of the $\Omega Z_2$-projection acting on
DD-bosons. The action on fermions can instead be read off from the
eigenvalues under $\Pi_p=\gamma^1....\gamma^{p-1}$, i.e. half plus
and half minus. Open string zero mode contributions come on the
other hand from bosonic zero modes $a_0^{I_N}$ along NN directions
and fermions  with a definite $\Pi_p$-chirality.

The amplitudes (\ref{direct}) define (by construction) a projection
of the closed and open string spectrum defined by the
oriented torus and
annulus amplitudes. However in order to state the consistency
of the string vacuum one should ensure that worldsheet boundaries
couple consistently to the bulk modes, i.e. that the sum of
the three amplitudes (\ref{direct}) in the closed
string channel reconstruct a complete square $(|B\rangle +
|C\rangle )^2$.
 Using the modular invariance of $Z(\tau)$ the amplitudes (\ref{direct})
can be rewritten in the closed string channel as
\bea
{\cal K} &=& {1\over 2}\int d\ell \, d\tilde{p}^+\,d\tilde{p}^-
e^{-\pi \alpha^\prime
\tilde{p}^+\tilde{p}^-}\, Z^{1\over 2}(i\ell)\,
\tilde{{\cal K}}_{0}\nn\\
{\cal A} &=& {1\over 8}\int d\ell \, d\tilde{p}^+\,d\tilde{p}^-
e^{-\pi \alpha^\prime
\tilde{p}^+\tilde{p}^-}\, Z^{1\over 2}(i\ell)\,
\tilde{{\cal A}}_{0}\nn\\
{\cal M} &=& {1\over 2}\int d\ell \, d\tilde{p}^+\,d\tilde{p}^-
e^{-\pi \alpha^\prime
\tilde{p}^+\tilde{p}^-}\, Z^{1\over 2}(i\ell+\ft12)\,
\tilde{{\cal M}}_{0}
\label{transverse}
\eea
 By $\ell$ we have denoted the uniforming length, given in terms of
the original variables in (\ref{direct}) via $\ell_K={1\over 2\tau_2}$,
 $\ell_A={2\over \tau_2}$, $\ell_M={1\over 2\tau_2}$.
Although it is not explicitly indicated, amplitudes in the
transverse channel are evaluated on masses $\tilde{\mu},\tilde{f}$
related to those in the direct channel by $S$ or $ST^2ST$ modular
transformations. More precisely $\tilde{\mu},\tilde{f}={\mu\over
\ell},{f\over \ell}$ for the Klein and Annulus amplitude and
$\tilde{\mu},\tilde{f}={\mu\over 2\ell},{f\over 2\ell}$ for the
Moebius. Finally $\tilde{p}^+\tilde{p}^-={p^+p^-\over \ell}$ for
the Klein and Moebius amplitude while
$\tilde{p}^+\tilde{p}^-={4p^+p^-\over \ell}$ for the annulus.
  Rewriting (\ref{transverse}) in terms of the transverse
variables we find:
\bea
\tilde{{\cal K}}_{0}&\equiv& {{\cal K}_0\over
Z_0^{1\over 2}(2i\tau_2)}=\prod_{i=1}^4 \left({1+e^{-\pi \tilde{\mu}_i}\over
1-e^{-\pi \tilde{\mu}_i}}\right)^{\epsilon_i}\nn\\
\tilde{{\cal A}}_{0}&\equiv& {{\cal A}_0\over
 Z_0^{1\over 2}({i\tau_2\over 2})}=
\prod_{i=1}^4 e^{-\pi \tilde{\mu}_i} \left(1-e^{-2\pi \tilde{\mu}_i}
\right)^{-\epsilon_i}\nn\\
\tilde{{\cal M}}_{0}&\equiv& {{\cal M}_0\over Z_0^{1\over
2}({i\tau_2\over 2}+{1\over 2})}= \prod_{i=1}^4 e^{-\ft12\pi
\tilde{\mu}_i} \left(1-e^{-\pi \tilde{\mu}_i}
\right)^{-\epsilon_i} \eea Remarkably after summing up $\ft12
{\cal K}_{0}+\ft{N^2}{8}\, {\cal A}_{0}\pm \ft{N}{2}\,{\cal
M}_{0}$ a complete square $(|B\rangle+|C\rangle)^2$ is
reconstructed \bea {1\over 2}\left[1\pm {N\over 2}\, \prod_{i=1}^4
e^{-\ft12 \pi \tilde{\mu}_i} (1+e^{-\pi
\tilde{\mu}_i})^{-\epsilon_i} \right]^2 \prod_{i=1}^4 \left({
1+e^{-\pi \tilde{\mu}_i}\over 1-e^{-\pi \tilde{\mu}_i}}
\right)^{\epsilon_i} \label{square} \eea For the D7-O7 system in
the D3-wave $\mu_i=f=\mu$ one recovers the results of
\cite{Hammou:2002bf}. It would be nice to give an interpretation
of the two terms in (\ref{square}) in terms of reflection
coefficients of closed string states in front of the
boundaries/crosscaps.
 Finally following \cite{Sinha:2002di,Hammou:2002bf},
we can derive tadpole conditions\footnote{Notice that amplitudes
(\ref{direct}) are in the odd-spin structure are therefore the
tadpole (\ref{square}) can be related to anomalies according to
\cite{Bianchi:2000de}.} by requiring that the combination inside
the square in (\ref{square}) cancels in the $\mu\rightarrow 0$
limit.
 This boils down to choose the orthogonal projection for the Moebius
and
\be
 N=2^{p-4}
\ee
 which matches the familiar result in flat space.

 In particular specifying to $p=9$ we conclude that 32 D9-branes are
need to cancel the O9 tadpole in the D1-wave.  Similarly the
charge of a O5-plane is canceled by 2 D5-branes sited at the
$R^4/Z_2$ singularity.

\section{Summary and open questions}

In this paper we applied exact CFT techniques to
solvable string models on RR/NSNS plane wave
geometries. Dp-waves descend from Penrose limits of
non-conformal Dp-brane geometries along critical null geodesics.
The solutions preserve sixteen supercharges and provide the
simplest examples of string vacua in the presence of non-self dual RR fluxes.
 We have focused on D1-plane waves of typeIIB/type I theory.
 These systems are natural candidates
to describe large ${\cal R}$-charge sectors of
 the type IIA/heterotic matrix string theories \cite{Dijkgraaf:1997vv}.
Before the
Penrose limit is taken they are described by warped $AdS_3\times S^7$
geometries and have been recently studied in \cite{Morales:2002ys}.
 Here we derive the bosonic KK spectrum of
${\cal N}=1,2$ supergravity on D1/F1 plane waves.
 KK particles organize nicely in groups of 256 (or 128 in type I
case) states at each floor of the KK tower.
  An important difference with the case of self-dual plane waves
is that excitations along or perpendicular to the RR-flux carry
different energies. This is due to the fact that the radii of the
$AdS_{p+2}$ and $S^{8-p}$ spaces in the parent warped geometry are
different.

 In section 4, we computed helicity traces counting multiplicities and charges
of string states in the pp-wave background.
  It would be nice to apply these helicity formulas to the study of
correlation functions in pp-wave backgrounds.
 As discussed earlier in that section, the helicity trace provides
an efficient description of the coupling of RR-plane waves to the
NSNS B-field. By expanding the resulting modular functions one
can derive one-loop correlation functions involving an arbitrary
number of $H=dB$ insertions. This can be used to test the
gauge/pp-wave correspondence beyond the planar level.

 Finally in subsection 4.2 we considered open descendants built from
D1-wave type IIB vacuum by adding Dp-Op branes
with $p=5,9$. Klein, Annulus and Moebius amplitudes are computed.
We test the consistency of the open string descendant by
showing that after modular transformations to the closed string
channel the three amplitudes combine themselves into a complete square.
Op-tadpoles are shown to be canceled by the introduction $2^{p-4}$
Dp-planes much in the same way as in flat space time.

   Our results show that Dp-waves (\ref{vacuum}) share most of the
nice integrability properties of the more familiar pp-waves
supported by self-dual fluxes. Besides the applications to
holography the study of these string vacua
is interesting by itself since they provide the simplest
 string solution with non-self-dual RR fluxes and constant dilaton.
 D1-waves of type IIB/I are related via S-duality to
pure NSNS F1-plane waves in type IIB/heterotic ten
dimensional string theories.
 Although RR-fluxes are present, the two sides of the duality
is described by a solvable string theory.
String dualities in the presence of RR-fluxes are fully unexplored.
The D1/F1-wave string dual pairs under consideration
here can provide a handle for this physics.

   Finally according to \cite{Gava:2002xb}, self-dual
D1-waves are related to blowing up deformations
of the dual $M^N/S_N$  SCFT's describing the
infrared regime of D1D5 systems on $M=T^4,K3$.
It would be nice to understand
whether these results extend to the 10D D1-waves studied
here. The candidate SCFT's are in this case given in terms of
second quantized type IIB/heterotic strings on $(R^8)^N/S_N$
\cite{Dijkgraaf:1997vv}.

\subsection*{Acknowledgements}

\noindent
We wish to thank A.B. Hammou, R. Russo, H. Samtleben,
M. Taylor, M. Trigiante and S. Vandoren for useful discussions.
 This work is partly supported by EU contract
HPRN-CT-2000-00122.

\mathon
\section*{Appendix: Partition functions and generalized modular forms}
\mathoff

In this appendix we compute, using the Lagrangian formalism,
 the one-loop partition function
of two dimensional massive fields coupled to a $U(1)$ current.
 We adopt the $\xi$-regularization formalism \cite{Ginsparg:1989pv}.

Let us start by considering a single complex boson with two-dimensional
action
\be
{\cal L}=\partial Z\,\bar{\partial} \bar{Z}-{\pi\, \bar{v}\over \tau_2}\,
\bar{Z}\partial Z-{\pi\, v\over \tau_2}\,\bar{Z}\bar{\partial}Z+
\pi^2 \left(\mu^2+{v \bar{v} \over \tau_2^2}\right)\,\bar{Z}\,Z+{\rm h.c.}
\label{lagrangian}
\ee
The coupling of the boson to the $U(1)$ current is described by the complex
number $v$, that will be written as $v=a\tau+b$.
After expanding in Fourier modes
\be
Z=\sum_{n_1,n_2}{} \, c_{n_1,n_2}\, e^{{\pi\over \tau_2}
\left[ n_2(z-\bar{z})
+n_1\,(\tau\,\bar{z}-\bar{\tau}\,z)\right]}
\label{determinant}
\ee
and performing the Gaussian integrals, it results in
the inverse of the determinant:
\bea
 {\rm det}\, \square_v &=&
\prod_{(n_1,n_2)} {\pi^2\over \tau_2^2}
\, \left(|n_2-n_1\tau-v|^2+\mu^2\tau_2^2\right)\nn\\
&=& \prod_{(n_1,n_2)} {\pi^2\over \tau_2^2} \,
|n_2-(n_1+a)\tau_1-b+i\tau_2\sqrt{(n_1+a)^2+\mu^2}|^2
\label{detprod} \eea Performing the $n_2$ infinite product by
means of the $\xi$-regularization formulas \be
\prod_{n=-\infty}^\infty \, (n-a)=2i\,\sin{\pi a}\quad\quad\quad
\prod_{n=-\infty}^\infty \, a=1 \ee we are finally left with \bea
 {\rm det} \, \square_v &=&
 \Theta_{a,b}(\mu|\tau,\bar{\tau})\equiv e^{4\pi \tau_2 \Delta_\mu}\,
 \prod_{(n_1 \in \,\bbZ+a)}\,\left|1-e^{-2\pi\tau_2\sqrt{\mu^2+n_1^2}
+2\pi i n_1 \tau_1 +2\pi i b} \right|^2 \label{theta} \eea where
$\Delta_\mu\equiv {1\over
2}\sum_{n=-\infty}^{\infty}\,\sqrt{\mu^2+n^2}$. In the massless
limit $\mu\rightarrow 0$ the generalized function (\ref{theta})
reduce to the familiar theta function: \be \lim_{\mu\rightarrow 0}
\Theta_{a,b}(\mu|\tau,\bar{\tau})= e^{-2\pi \tau_2 a^2}
\left|\vartheta_1(a\tau+b|\tau)\over \eta \right|^2 \ee The
modular invariance of this form can be read from (\ref{detprod}),
which is clearly invariant under\footnote{In proving this one uses
$\prod_{n=-\infty}^\infty\tau_2=1$.}: \be \tau\rightarrow -{1\over
\tau}\quad\quad \quad\quad \mu\rightarrow \mu |\tau| \quad\quad
v\rightarrow -{v\over \tau} \quad\quad n_1\leftrightarrow n_2.
\label{modular} \ee According to $v'=a'\tau'+b'=-{v\over \tau}$,
we see that spin structures transform under the modular group as
$a'=b,b'=-a$.

Similar manipulations hold for the fermionic determinants with
the result (\ref{theta}) now coming in the numerator.

\end{document}